\documentclass[%
 reprint,
 amsmath,amssymb,
aps,
prl,
floatfix,
]{revtex4-1}

\usepackage{graphicx}
\usepackage{amsmath}	
\usepackage{amssymb}	
\usepackage{dcolumn}
\usepackage{color}
\usepackage{bm}
\usepackage{newtxtext,newtxmath}
\usepackage[T1]{fontenc}
\usepackage{ae,aecompl}
\usepackage{natbib}
\usepackage{blindtext}
\usepackage{aas}
\usepackage{hyperref}

\begin{document}

\preprint{APS/123-QED}

\title{A local fading accelerator and the origin of TeV cosmic ray electrons}

\author{S. Recchia}
\email[E-mail: ]{recchia@apc.in2p3.fr}
\author{S. Gabici}
\affiliation{APC, AstroParticule et Cosmologie, Universit\'e Paris Diderot, CNRS/IN2P3, CEA/Irfu, Observatoire de Paris, Sorbonne Paris Cit\'e,\\10, rue Alice Domon et L\'eonie Duquet, 75205 Paris Cedex 13, France}
\author{F. A. Aharonian}
\affiliation{Dublin Institute for Advanced Studies, 31 Fitzwilliam Place, Dublin 2, Ireland}
\affiliation{Max-Planck-Institut f\"ur Kernphysik, Postfach 103980, D-69029 Heidelberg, Germany}
\author{J. Vink}
\affiliation{GRAPPA, University of Amsterdam, Science Park 904, 1098 XH Amsterdam, Netherlands}
\affiliation{API, University of Amsterdam, Science Park 904, 1098 XH Amsterdam, Netherlands}
\affiliation{SRON Netherlands Institute for Space Research, Utrecht, The Netherlands}
\date{\today}

\begin{abstract}
The total cosmic ray electron spectrum (electrons plus positrons) exhibits a break at a particle energy of $\sim 1\rm~TeV$ and extends without any attenuation up to $\rm \sim 20~ TeV $.
Synchrotron and inverse Compton energy losses strongly constrain both the age and the distance of the potential sources of TeV and multi-TeV electrons to $\rm\approx 10^5~yr$ and $\rm \approx 100-500~pc$, depending on both the absolute value and energy dependence of the cosmic ray diffusion coefficient.
This suggests that only a few, or  just one nearby discrete source may explain the observed spectrum of high energy electrons. 
On the other hand the measured positron fraction, after initially increasing with particle energy,  saturates at a level well below 0.5 and likely drops above $\sim 400-500$ GeV.
This means that the local source(s) of TeV electrons should not produce positrons in equal amount, ruling out scenarios involving pulsars/pulsar winds
as the main sources of high energy leptons.
In this paper we show that a single, local, and fading source can naturally account for the entire spectrum of cosmic ray electrons in the TeV domain.
Even though the nature of such source remains unclear, we discuss known cosmic ray accelerators, such as supernova remnant and stellar wind shocks, which are believed to accelerate preferentially electrons rather than positrons.
\end{abstract}

\maketitle


\section{Introduction}
\label{sec:intro}

Recently the HESS \cite{HESS-2017-HEE}, DAMPE \cite{DAMPE-2017-ele} and CALET \cite{CALET-2017-ele-pos} collaborations 
reported on the presence of a break  at $\sim 1\rm~TeV$ in the cosmic ray (CR) electron spectrum. 
Moreover, the HESS collaboration  measured the electron spectrum up to $\rm \sim~20~TeV$ \citep{HESS-2017-HEE}  and found, remarkably, that it does not exhibit any cut-off up to this very high energy.

The severe energy losses suffered by high energy (multi-TeV) leptons in the interstellar medium (ISM), mainly due to Compton and synchrotron cooling, during their diffusive motion from their sources to Earth, limit both the age and the distance of the potential electron TeVatrons to $\rm\approx 10^5~yr$ and $\rm \approx 100-500~pc$, for accepted values of the CR diffusion coefficient (see e.g \cite{Atoyan-1995-long, Manconi-2018-HEE}).
Given these very strict limitations on age and distance, most likely very few, or possibly only one local astrophysical source may account for the TeV electron flux, suggesting that a single local source approach is very plausible at TeV energies, instead of a multiple far-away sources description, which is suitable at lower energies, where energy losses are less severe.

Another important piece of information comes from the measured positron fraction (\cite{Pamela-2013-pos-frac, FERMI-2012-ele-pos, AMS02-2014-pos-frac}). This quantity is observed to grow in the energy range $\rm \sim 10-200~GeV$, a trend that cannot be reproduced by secondary production of electrons and positrons in CR interactions in the ISM. Thus additional sources of electrons/positrons are needed (see e.g \cite{Aharonian-1991-optical-UV, Dogiel-1990,Hooper-2009-pulsar,Blasi-2009-pos-excess, Profumo-2012-review, Atoyan-1995-long, Atoyan-1995-short}). Moreover, the observed positron fraction flattens to a value of $\sim 0.15$  at $\sim 200\rm~GeV$ \citep{AMS02-2014-pos-frac}, and likely drops above $\gtrsim \rm 400-500~GeV$ \citep{AMS02-2018-pos}. 
This limits the potential contributors to the TeV leptonic flux to sources that accelerate mainly electrons, such as supernova remnants (SNR) or stellar winds, while scenarios that involve the production of an equal amount of electrons and positrons up to multi-TeV energies, such as pulsars/pulsar winds (see e.g \cite{Giacinti-2018}),  
are strongly disfavored.

In this paper we explore a single, local electron source scenario for the interpretation of the observed multi-TeV CR electron spectrum. 
We adopt a setup  similar to the one proposed in \cite{Atoyan-1995-long}, and we consider separately  the flux from distant sources (beyond $\rm \sim~500~pc$ from Earth), modeled as a continuous, stationary and homogeneous distribution, and the flux from a single nearby point source, whose injection is modeled either as a burst, or as a continuous injection over an extended time and with a luminosity that decreases with time (i.e. a fading source).

Note that this is a correct mathematical calculation of the flux  from  discreetly  distributed sources. In fact, if  the characteristic distance between the sources, $d$, (as well the mean free path of particles) is large, one cannot estimate the total flux as in the case of a continuous distribution of sources, namely by integrating from zero (the position of the observer) to infinity. Instead, one should consider separately the contribution   between $d$ and infinity (which can be approximated as a continuous source distribution) and the one from the nearby sources (see e.g \cite{Atoyan-1995-long}).

The injection spectrum and maximum energy of the two components are considered as parameters, although we assume that both the components (distant sources and local one) belong to the same population, so that their injection parameters are not expected to be too different. Notice that the above mentioned sources are taken to produce electrons only.

The propagation of leptons from the sources is then treated in the diffusive regime, with the energy-dependent diffusion coefficient $D = D_0 ~(E/{\rm TeV})^{\delta}$ taken as a parameter, with both normalization and slope compatible with the measurements of the B/C ratio, and also taking into account energy losses,  due in particular to  Compton and synchrotron cooling.

We show that the electron  flux for energies below $\rm \sim~100-300~GeV$ can be accounted for by the distant sources if their injection spectrum has a slope $\alpha \sim 2.3-2.5$. At higher energies, instead, the relative importance of the distant and local sources depends on the normalization and energy dependence of the diffusion coefficient.

Below, we demonstrate that the TeV break in the CR electron spectrum can be naturally interpreted as a cooling break, corresponding to the energy at which the age of the source equals the loss time in the ISM. Moreover, we show that the reported fluxes  up to 20 TeV could be naturally explained by a single operating electron TeVatron in its fading stage. Our conclusion is that no room is left for interpretations based on sources that produces electrons and positrons in equal amount \cite{Giacinti-2018}, and we briefly discuss the possible role played by known sources that preferentially accelerate CR electrons rather than positrons, such as supernova remnants and stellar winds.

\section{A single source scenario}
\label{sec:single-src}

The propagation of CR electrons in the ISM is described by the transport equation (e.g. \cite{Berezinskii-1990-bible})
\begin{equation}\label{eq:trasport}
\frac{\partial f(t, \vec{r}, E)}{\partial t} -D(E) \vec{\nabla}^2f(t, \vec{r}, E) + \frac{\partial }{\partial E}\left(b(E)f(t, \vec{r}, E)\right)=Q(t, \vec{r}, E),
\end{equation}
where $f(t, \vec{r}, E)$ is the electron distribution function, $E$ is the particle energy, $D(E)$ is the diffusion coefficient, $b(E)$ is the energy loss rate and $Q(t, \vec{r}, E)$ is the injection spectrum.

In what  follows we assume that the diffusion coefficient  is spatially uniform, isotropic, charge independent, and depends on the particle energy as $D = D_0 ~(E/{\rm TeV})^{\delta}$, where $\delta$ is expected to assume values in the range $\sim 0.3-0.6$ (\cite{Strong-2007-review,Galprop-2011-propa}). During the propagation in the ISM, electrons lose energy via ionization/Coulomb, Bremsstrahlung, and synchrotron/inverse Compton interactions at low, intermediate, and high particle energies, respectively, at a rate \citep{Atoyan-1995-long}:
\begin{equation}
\label{eq:losses}
\frac{{\rm d}E}{{\rm d}t} \sim a_i+a_B \left( \frac{E}{\rm TeV} \right) + a_{s/C} \left( \frac{E}{\rm TeV} \right)^2,
\end{equation}
where $a_i \approx 10^{-7} (n/{\rm cm}^{-3})$ eV/s describes the ionization losses in a ISM of density $n$, $a_B \approx 7 \times 10^{-4} (n/{\rm cm}^{-3})$ eV/s the Bremsstrahlung energy losses, and $a_{s/C} \approx 0.1 ~(w/{\rm eV~cm}^{-3})$ eV/s the synchrotron and inverse Compton energy losses, with $w$ representing the sum of the energy density of soft ambient photons and of the Galactic magnetic field. Note that typical values in the ISM are $n=1~ \rm cm^{-3}$ and $w = 1~\rm eV/cm^3$. Ionization losses dominates for particle energies in the sub-GeV domain, synchrotron and inverse Compton losses dominates for particle energies above the multi-GeV domain, and Bremsstrahlung dominates in the intermediate energy range.

Consider a source of CRs located at a distance $d$ from the observer which emits continuously electrons. Here we assume that the source begins to release electrons in the ISM at time zero, so that its age $t_{a}$ is equal to the present time. 
The observed electron spectrum extends up to particle energies of $\rm E_{max}^{obs} \sim 20$ TeV (see the H.E.S.S. data points in Fig.~\ref{fig:1}), without showing any cutoff.
Obvious necessary conditions to receive  particles of such energy from the source are that both the source age $t_{a}$ and the energy loss time $t_l(20~{\rm TeV}) \sim 2 \times 10^4 (w/{\rm eV~cm}^{-3})^{-1}$ yr (for particle energies of $\sim$ 20 TeV synchrotron and inverse Compton losses dominate) should be larger than the diffusion time $t_d = d^2/6~D(20~{\rm TeV})$.
The latter condition provides a constraint on the distance of the source:
\begin{equation}
\left( \frac{d}{100~{\rm pc}} \right) 
\lesssim ~ \left( \frac{D_0}{10^{29}~{\rm cm^2/s}} \right)^{1/2} \left( \frac{w}{\rm eV/cm^3} \right)^{1/2}.
\end{equation}

In the next Sections we investigate the possibility that the same source of $\sim$20 TeV electrons is also responsible for the appearance of the spectral break reported by HESS and DAMPE at  $E_{br} \sim 1~\rm TeV$.
If we impose that the TeV feature is due to cooling in the ISM, we can immediately derive an estimate of the source age by equating it to the energy loss time: $t_a = t_l(E_{br})$, which gives:
\begin{equation}
\label{eq:age}
t_a \sim 3 \times 10^5 \left( \frac{w}{\rm eV/cm^3} \right)^{-1} {\rm yr}
\end{equation}

In the following, we will show that a source that releases continuously CR electrons from time $t = 0$ up to the present time ($t = t_a$) could explain, under certain conditions, the  electron spectrum above $\sim 1$ TeV, while a burst-like source would fail to do so.


\subsection{Burst-like source}
A burst-like source of CR electrons located at position $\vec{r}_s=0$ is characterized by an injection rate of particles of the form:
\begin{equation}
Q(t, \vec{r}, E)=S(E)\delta(t)\delta(\vec{r}).
\end{equation}
For simplicity, we limit ourselves to considering power law particle spectra: $S(E) = S_0 (E/{\rm TeV})^{-\alpha}$.
The solution of the transport equation (Eq.~\ref{eq:trasport}) is \cite{Atoyan-1995-long}:
\begin{align}\label{eq:point-burst}
& f(t_a,\vec{r}, E) = \frac{S(E_{t_a})}{\pi^{3/2}r_d^3(E, E_{t_a})} \frac{b(E_{t_a})}{b(E)} \exp\left[{-\frac{\vec{r}^2}{r_d^2(E, E_{t_a})}}\right]\\
&t_a =\int_E^{E_{t_a}}\frac{dE'}{b(E')} \approx \frac{E~E_{br}(t_a)}{E + E_{br}(t_a)}\nonumber \\
&r_d^2(E, E_{t_a})\equiv 4\int_E^{E_{t_a}}\frac{D(E')}{b(E')}dE' \approx 4~D(E)~t_E,\nonumber
\end{align}
where the approximations have been obtained for the energies where inverse Compton and  synchrotron energy losses dominate. 
Under this approximation, the break energy is $E_{br}(t_a)=\frac{1}{a_{s/C}t_a} $, and $t_E= min\left(t_a, \frac{1}{a_{s/C}E}\right)$.
Particles of energy $E_{t_a}$ cool down to $E$ during a time $t_a$. $r_d$ is the diffusion length. 
Notice that $f(t_a,\vec{r}, E) $ can be nonzero only if $E_{t_a}(E) > E$, and if $E_{t_a}(E) < \rm E_{max}$, i.e the maximum energy of the injection spectrum. 
An example of point burst-like source solution is shown in  Fig.~\ref{fig:1}. The age of the source has been chosen so that $t_a = \frac{1}{a_{s/C}E_{br}}$ corresponds to $E_{br} \sim 1$ TeV. This gives $t_a\approx 10^5~\rm yr$. 

For particle energies smaller than $\sim 0.5-0.75 ~E_{br}$ energy losses can be neglected. Moreover, for energies larger than
\begin{equation}
E_{diff} = \left[5\times 10^{-2} \left(\frac{r}{\rm 100~pc}\right)^2 \left(\frac{t_a}{10^5\rm~yr}\right)^2 \left(\frac{D_0}{10^{29}\rm~cm^2/s}\right)^2  \right]^{\frac{1}{\delta}} \rm TeV, 
\end{equation}
the diffusion time is shorter than $t_a$, namely the diffusion length is larger than $r$ and the exponential in Eq.~\ref{eq:point-burst} can be neglected. For typical values of the parameters, $E_{diff}\approx$ few GeV.  
In this regime the electron spectrum can be approximated as
\begin{equation}
\label{eq:fburst}
f(t_a,\vec{r},E) \sim \frac{S(E)}{\pi^{3/2} r_d^3} \propto E^{-\alpha-3/2\delta}.
\end{equation}
It is interesting to note that due to propagation the spectrum is steepened by a power $3/2\delta$ with respect to the injected one.
For energies below $E_{diff}$ the solution exhibits an exponential suppression. 

For energies larger than $E_{br}$ the electron cooling time is shorter than the age of the source and thus particles of these energies cannot reach the observer.
This implies that a very sharp cutoff appears in the spectrum, so that it is impossible to reproduce with a single burst-like source both the spectral feature at $E \sim 1$ TeV and the extension of the spectrum up to $\sim$ 20 TeV.
Despite this fact, it is still useful for the remainder of our discussion to estimate the total energy that the burst-like source has to inject in form of CRs in order to reproduce at least the observed electron spectrum at particle energies of the order of $E = E_{br} \sim 1$ TeV.
By fitting $f(t_a,E)$ (from Eq.~\ref{eq:fburst}) to the DAMPE data ($\approx 4 \times 10^{-6}$ eV/cm$^3$) at $E = E_{br}$  we obtain:
\begin{equation}
S_0 E_{br}^2 \approx 5 \times 10^{47} {\rm erg}
\end{equation}
which for $\alpha = 2$ corresponds to a total energy input in CR electrons equal to:
\begin{equation}\label{eq:energy-burst}
W_{b} \approx 7 \times 10^{48} {\rm erg}
\end{equation}
in the range of particle energies spanning from $E = m_e c^2$ (electron rest mass energy) to $E \sim 1$ TeV. Note that this value is somewhat larger than that expected for electrons from SNRs.


\subsection{Continuous point source}
Here we assume that a source, located  at $\vec{r}_s=0$, turned on at time $0$ and continuously injected electrons with a power-law spectrum $S(E) = S_0 (E/{\rm TeV})^{-\alpha}$ up to the present time $t_a$. The source luminosity $L(t)$ is assumed to decrease  with time.
The  injection term is given by
\begin{equation}
 Q(t, \vec{r}, E)=S(E)L(t)H(t_a)\delta(\vec{r}),
\end{equation}
so that the solution of the transport equation, (Eq.~\ref{eq:trasport}), reads \cite{Atoyan-1995-long}:
\begin{equation}\label{eq:point-cont}
f(t_a,\vec{r}, E) =\int_0^{t_a} dt' \frac{S(E_{t'})L(t')}{\pi^{3/2}r_d^3(E,E_{t'})} \frac{b(E_{t'})}{b(E)} e^{-\frac{\vec{r}^2}{r_d^2(E, E_{t'})}}.
\end{equation}
In the case of a constant luminosity, and taking into account the approximations used in Eq.~\ref{eq:point-burst}, the resulting electron spectrum is given by:
\begin{equation}
f(t_a, \vec{r}, E) = \frac{L_0 S(E)}{4\pi~D(E)~r} {\rm erfc}\left(\frac{r}{2\sqrt{D(E)t_E}}\right).
\end{equation}
Note that for energies such that $r<2\sqrt{D(E)t_E}$ the energy dependence of the solution is
\begin{equation}
f(t_a, \vec{r}, E)\propto E^{-\alpha-\delta},
\end{equation}
where the steepening due to propagation is by a power $\delta$, in contrast with the $3/2\delta$ dependence in case of  burst injection (Eq.~\ref{eq:fburst}).

\begin{figure}
	\centering
	\includegraphics[width=\columnwidth]{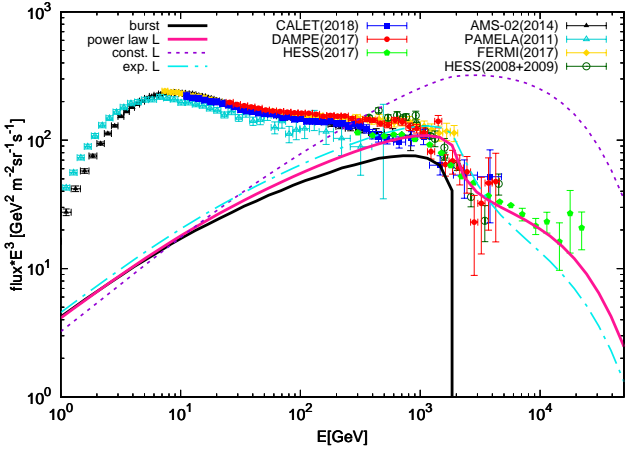}
    \caption{Examples of electron spectra from a single point source, in the case of different types of injection: burst-like (black solid line), $L(t)=
L_0/\left[1+\frac{t}{\tau}\right]^2$ (magenta solid line, $\tau/t_a \sim 0.05$), constant luminosity (purple dashed line),  $L(t)=L_0e^{-t/\tau}$ (cyan dot-dashed line, $\tau/t_a \sim 0.2$). In all cases: $\rm D(E)=10^{28}(E/10~GeV)^{0.3}~cm^2/s$, $\alpha=2.3$, $\rm t_a=10^5~yr$, $\rm d=100~pc$. Data points are from \cite{AMS02-2014-ele,CALET-2017-ele-pos,FERMI-2017-ele-pos,HESS-2017-HEE,DAMPE-2017-ele, PAMELA-2011-ele, HESS-2008-ele-pos, HESS-2009-ele-pos}.}
	\label{fig:1}
\end{figure}

The case of a fading source is somewhat intermediate between that of a burst-like source and of a constant luminosity source. 
Let us assume, for instance, a time dependence of the luminosity in the form
\begin{equation}
\label{eq:luminosity}
L(t)=\frac{L_0}{\left[1+\frac{t}{\tau}\right]^{\gamma}},
\end{equation}
with $\gamma>0$. 
If $\tau/t_a$ is small ($\lesssim 0.01$) or $\gamma$ is large ($\gtrsim 3$), the source releases most of electrons in a short time compared to the source age and the solution will be similar to that of a burst-like source. In the opposite case, the solution will resemble more to the case of constant luminosity. The combination of these two behaviors is shown in Fig.~\ref{fig:1}, where a case with $\gamma=2$ and $\tau/t_a \sim 0.05$ is shown with a solid magenta line. The spectrum presents a drop at energy $\approx E_c(t_a)=\frac{1}{a_{s/C}t_a}$, where a burst-like source of the same age presents a sharp cut-off due to cooling. 
The amplitude of the depth depends on
$\tau/t_a$ and is large (small) if this ratio is small (large). 
Note however that the functional form in Eq.~\ref{eq:luminosity} does not play any special role, and the same considerations apply to  other temporal dependences of the luminosity. For instance, in Fig.~\ref{fig:1} we also show the case of $L(t)=L_0e^{-t/\tau}$ with  $\tau/t_a \sim 0.2$ (cyan dot-dashed line).

Remarkably, the electron spectrum above $\sim 1~\rm TeV$ can be explained fairly well with a single continuous fading source. This was not possible with a single burst-like source.

\subsection{Continuous and stationary injection in a disk}
In order to model the contribution of distant sources to the electron spectrum we consider a continuous, stationary and homogeneous distribution of sources on a disk of radius $R$ and hight $h$, injecting a spectrum $S(E)$ beyond a given distance $r_0$ from Earth, assumed to be located at $\vec{r}=0$.
The corresponding injection term reads
\begin{equation}
 Q(t, \vec{r}, E)=S(E)\exp\left(-\frac{x^2 + y^2}{R^2}-\frac{z^2}{h^2}\right) ~~~~~ {\rm for} ~~ r > r_0
\end{equation}
which, substituted in Eq.~(\ref{eq:trasport}), gives
\begin{equation}\label{eq:inj-disk}
f(\vec{r}=0, E) =\int_E^{\infty}dE' \frac{S(E')}{2b(E)}\frac{e^{-\frac{r_0^2}{r_d^2}\left( 1+\frac{r_d^2}{R^2}\right)}}{\left( 1+\frac{r_d^2}{R^2}\right) \sqrt{1+\frac{r_d^2}{h^2}}}.
\end{equation}

\begin{figure}
	\centering
	\includegraphics[width=\columnwidth]{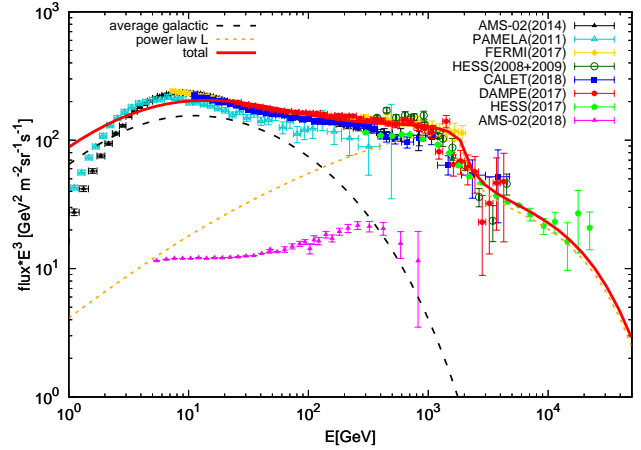}
    \caption{Possible fit (red solid line) to the total observed electron spectrum  in the case of $\rm D(E)=10^{28}(E/10~GeV)^{0.3}~cm^2/s$, due to: distant (beyond $\rm 500~pc$) sources (black dashed line, $\alpha=2.4$); a local continuous fading source (orange dotted line, $\alpha=2.3$, $\rm d=100~pc$, $\rm t_a=10^5~yr$, $\rm \tau/t_a=0.08$). The positron flux is also shown (magenta points) \cite{AMS02-2018-pos}.}
	\label{fig:2}
\end{figure}

\begin{figure}
	\centering
	\includegraphics[width=\columnwidth]{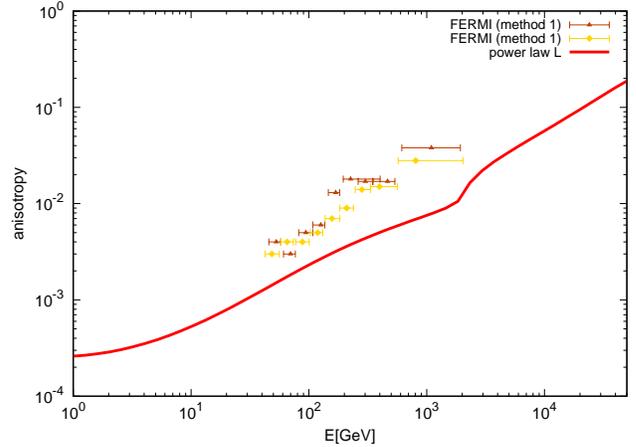}
    \caption{Dipole anisotropy estimated for the continuous source (red line) used for the fit of Fig.~\ref{fig:2}.}
	\label{fig:3}
\end{figure}

\section{Fit to the data}
\label{sec:fit}
In the previous Section we have shown that a single continuous source can explain the electron spectrum above $\sim 1$  TeV. This is visible in Fig.~\ref{fig:2}, where we present a possible fit to the electron spectrum. The fit has been obtained by assuming that the distant sources are uniformly distributed on a disk of radius and height ($\rm R= 15~kpc$, $\rm h = 150~pc$) beyond $\rm r_0=500~pc$ from the observer and inject a spectrum with slope and maximum energy ($\rm \alpha = 2.4$, $\rm  E _{max} \gtrsim 10^5~GeV$). The luminosity of the  source distribution  is $\rm \approx 1.5\times 10^{39}~erg/s$, which corresponds to an injection rate in electrons between $1.5-5\%$ of the total CR injection rate in the Galaxy.
The continuous source (Eq.~\ref{eq:luminosity}) is characterized by $\rm t_a\sim 10^5~yr$, $\rm d=100~pc$, $\rm \tau/t_a\sim 0.08$, $\gamma=2$, $\alpha=2.3$, and total energy $\rm \approx 4\times 10^{47}~erg$).
Such energy input corresponds to $\sim 0.4\%$ of the total energy in CRs expected to be injected by a SNR ($\rm \approx 10^{50}~erg$).
Notice that this energy requirement matches quite well that expected for electrons from SNRs, contrary to the one estimated for a burst-like source (Eq.~ \ref{eq:energy-burst}).
Moreover, if one assumes that the local source produces also CR protons with a total energy 100 times larger than that in electrons, namely $4\times 10^{49}$ erg, the  spectrum of the local source  would be subdominant compared to the observed one (see also  Fig. 2 of  \cite{Atoyan-1995-short}).

The diffusion coefficient is assumed as $\rm D(10~GeV) \sim 10^{28}~cm^2/s$ and $\rm \delta=0.3$, but larger slopes  (up to $\sim 0.6$) for the diffusion coefficient are also accepted \cite{Galprop-2011-propa}.

In the present scenario, distant sources largely dominate the electron spectrum at energies below $\approx 100$ GeV. Above that energy this contribution drops and the local source starts to dominate. Above $\approx 1$ TeV a very good fit to data is obtained.
In fact, a good fit to the data can be achieved for a large variety of parameters, such as the injection spectrum ($\alpha$ in the range $\approx 1.8 ... 2.4$), the maximum energy of the distant and local sources and the diffusion coefficient ($\delta \approx 0.3 ... 0.6$).
The functional form chosen to describe the fading of the source (Eq.~\ref{eq:luminosity}) provides a good fit to data for parameters in the range $2< \gamma <3$ and $\tau/t_a \approx 0.01 ... 0.1$. A lower value of $\gamma$ requires a correspondingly lower value of $\tau/t_a$.

In the scenario considered here the dipole anisotropy in the arrival direction of CR electrons is determined by the local source as:
$
a=(3D(E)/c) \vert \nabla~f_s\vert/f_{tot}
$
where $f_s$ is the flux corresponding to the local source and $f_{tot}$ is the total electron flux \cite{Berezinskii-1990-bible}. In Fig.~\ref{fig:3} we compare this estimate with the upper limits provided by Fermi \cite{FERMI-2017-ele-ani}, showing that the anisotropy is below the upper limits.

\section{Discussion and conclusions}
\label{sec:conclusions}

We have shown as a single, local, and fading accelerator of electrons can explain the entire high energy CR electron spectrum. Scenarios involving sources accelerating electrons and positrons in equal amount (as in standard model for pulsars/pulsar winds) are ruled out by data.

A crucial issue is to determine the nature of the local accelerator of electrons. Even though at the moment it is not possible to reach any firm conclusion, it is worth examining the possible role of known sources of CRs that are believed to accelerate preferentially electrons over positrons. These include, for example, supernova remnants and stellar winds. 

For the choice of parameters done in Fig.~\ref{fig:2} (injection spectrum $E^{-2.3}$, and a typical Galactic diffusion coefficient with $\delta = 0.3$), the fading source should convert $\sim 4 \times 10^{47}$ erg in CR electrons. 
SNRs can naturally match this energy requirements  (the total explosion energy is $\sim 10^{51}$ erg), and possibly also powerful winds from massive Wolf-Rayet stars \cite{Crowther-2007}.

What remains unclear is whether such sources could provide the temporal fading pattern required to fit the electron spectrum. 
We note that  the reference parameters used in Fig.~\ref{fig:2} describe a source of age $t_a = 10^5$ yr, which was active for an initial period of many thousands of years ($\tau/t_a = 0.08$), followed by a fading phase. $\tau$ is definitely shorter than the duration of the Sedov phase of a supernova remnant, and it is plausible to envisage the acceleration of multi-TeV electrons over such a time scale \cite{Lagage-1983}. The properties of the following fading phase would depend on the details of the escape process from the accelerator, and from the surrounding of the accelerator, which unfortunately are not firmly established (see e.g. \cite{Gabici-2011}). Justifying a fading temporal pattern for a stellar wind could  be even more difficult, given the poorer knowledge we have about this object as particle accelerators.

One may wonder why we do not see the electron source, if it is so close to us. A $10^5$ yr old supernova remnant would be in the radiative phase, most likely quite close to the end of its life. With a radius of few tens of parsecs, and at a distance of a hundred parsecs or so, could be missed because of its extremely large angular size (a diameter of about 10 degrees for a remnant radius of 20 pc located at 100 pc from us).

Finally, potential accelerators that could meet the age and  distance requirements are the old supernova remnant known as the Monogem Ring \cite{MonogemRing-2018} or the Wolf-Rayet star binary $\gamma^2$ Velorum (see e.g \cite{Sushch-2011}).
However, at this stage we do not want to be too specific, as there may be other potential SNRs or stellar wind accelerators as well.

\section*{Acknowledgements}
SR and SG acknowledge support from the region \^{I}le-de-France under the DIM-ACAV programme, from the Agence Nationale de la Recherche (grant ANR- 17-CE31-0014), and from the Observatory of Paris (Action F\'ed\'eratrice CTA).



\bibliographystyle{apsrev4-1}
\bibliography{biblio} 


\end{document}